# Tensile deformation and fracture mechanisms of Cu/Nb nanolaminates studied by *in situ* TEM mechanical tests

Z. Liu[a,b], M. A. Monclús[a], L. W. Yang[a], M. Castillo-Rodríguez[a], J. M. Molina-Aldareguía[a,*], J. Llorca[a,c,*]

[a]IMDEA Materials Institute, C/Eric Kandel 2, 28906, Getafe, Madrid, Spain

[b]College of Mechanical and Electrical Engineering, Central South University, 410083, P.R. China

[c]Department of Materials Science, Polytechnic University of Madrid, E.T.S. de Ingenieros de Caminos, 28040 Madrid, Spain

**Abstract**

The mechanisms of deformation and failure of Cu/Nb nanolaminates manufactured by accumulated roll bonding were analysed using *in-situ* TEM mechanical tests. Dog-bone small-scale tensile specimens were prepared using a FIB-based technique with the layers parallel and perpendicular to the loading axis. Load, deformation and TEM micrographs were recorded during the *in-situ* TEM mechanical tests. The specimens deformed parallel to the layers presented very high strain hardening and ductility while those deformed in the perpendicular orientation were more brittle. These results were rationalized in terms of the deformation and fracture mechanisms observed during the tests.

\* Corresponding authors: jon.molina@imdea.org (J. M. Molina-Aldareguía), javier.llorca@imdea.org (J. LLorca)

# 1. Introduction

Metallic nanolaminates (MNLs) are formed by alternating layers of different metals with layer thickness in the nm range (< 100 nm). Plastic deformation in the metallic layers is constrained by the high density of interfaces, leading to a very large strength and indentation hardness [1, 2], together with good thermal stability [3, 4], and promising fatigue/failure resistance [5]. Obviously, the mechanical properties of MNLs depend on the lattice structure of the metallic layers (fcc, bcc or hcp), the layer thickness and the characteristics of the interfaces (coherent, semi-coherent or incoherent). Many different MNLs have been studied in recent years, such as Al/Nb [6], Ag/Fe [7], Zr/Nb [8], Mg/Ti [9], V/Ag [10] and Cu/X with X=Nb [11], Zr [12], Ni [13], Ag [14], Cr [15], W [16], and Ru [17]. They were manufactured by magnetron sputtering because it is relatively easy to produce different MNLs with well-controlled layer thicknesses from a few nm to several tens of nm. The mechanical properties of these thin films have been studied by means of nanoindentation and micropillar compression tests, and the information obtained from these investigations has been used to determine the dominant deformation mechanisms [1, 12]. It is nowadays widely accepted that the flow strength of MNLs with thick layer thicknesses ($h$) > 50 - 100 nm is controlled by the formation of dislocation pile-ups at the interfaces [18, 19]. This mechanism is inhibited for smaller layer thicknesses (5 - 10 nm < $h$ < 50 - 100 nm) because of the lack of space to form the dislocation pile-ups and confined layer slip (CLS) becomes the dominant process [20-21]. The flow stress in this regime is proportional to $\ln(h/b)/h$ where $b$ is the Burgers vector and this mechanism has been validated by means of *in-situ* mechanical tests within a transmission electron microscope (TEM) [22]. For very small layer thickness ($h$ < 5 -10 nm), the stress necessary to move dislocations by CSL is higher than the stress necessary to drive the dislocations across the interface, which becomes the dominant deformation mechanism. In this regime, the flow stress becomes independent of the layer thickness (or decays slightly with decreasing the layer thickness) [21, 23]. The actual values of the flow strengths and layer thicknesses that lead to the transition among the different mechanisms depend on the interface characteristics and the type of metals in the MNL.

One important development from the viewpoint of engineering applications has been the manufacturing of MNLs by means of accumulated roll bonding (ARB) [24]. This method involves repeated rolling, sectioning, stacking, bonding, and re-rolling of the laminate, leading to a progressive reduction of the layer thickness down to the nm scale. The main advantage of ARB is that it allows the production of large sheets of MNLs with much larger total thicknesses than those obtained by magnetron sputtering, opening the way for the application of bulk MNLs in structural components. However, not all metallic materials are amenable to be processed into nanolaminates by ARB because it is not always possible to thin down the layers to the nm scale without breaking the continuity due to the onset of strain localization in the layers upon severe plastic deformation. Successful examples of MNLs manufactured by ARB include Cu/Nb, Cu/Ta and Cu/V [3, 19, 25-29]. Bulk samples of MNLs manufactured by ARB have been tested in tension parallel to the layers [19, 27] to assess the effect of layer thickness on the strength and ductility, which are important parameters from the viewpoint of engineering applications. Bulk tensile testing perpendicular to the layers was, however, not possible due to the small thickness of the ARB sheets. It was found that the strength increased as the layer thickness decreased [19, 27] and that the MNLs deformed in the transverse direction (TD) were stronger and presented higher strain-to-failure than those tested in the rolling direction (RD) [19]. However, the actual deformation and fracture mechanisms responsible for this behaviour were unknown because they cannot be ascertained in the fractured samples due to the nm scale of the relevant microstructural dimensions.

In order to ascertain the deformation and fracture processes of MNLs, mechanical tests within a TEM have been carried out in Cu/Nb [11, 22], Al/Nb [6], and Cu/Ni [13] MNLs processed by magnetron sputtering. Because the MNLs were in the form of thin coatings, nanoindentation [11, 22] or micropillar compression [6] were used to deform the MNLs. These tests provided evidence of deformation by CLS between layers as well as of dislocation climb at the interfaces, depending on the layer thickness and the interface strength. Moreover, fracture tests within a TEM have also been used to analyse the influence of the layer thickness on the weakest crack path during crack propagation in metal-ceramic nanolaminates [30, 31].

In this investigation, the deformation and fracture mechanisms of ARB Cu/Nb MNLs were analysed at the nm scale by means of tensile tests within a TEM. Samples were tested in two different orientations (parallel and perpendicular to the layers) and the load and displacement were continuously recorded in real time during the tests. They provided a very detailed picture of the deformation mechanisms as well as of the initiation and progress of fracture during tensile deformation as a function of orientation in these materials. This information addresses three fundamental issues, including (a) the fracture mechanisms of MNLs when deformed in different orientations, (b) the crack nucleation, propagation and arrest between different nanolayers, and (c) the effect of constrained plasticity on crack propagation. This information is important to understand and design MNLs with optimized tensile properties in terms of strength, ductility and toughness.

## 2. Material and experimental techniques

Cu/Nb MNLs were processed by ARB following the procedure described in [24, 29]. The initial Cu/Nb stack consists of alternating Cu (99.99% purity) and Nb (99.94% purity) layers of equal thickness. In order to ensure that the ARB microstructures were free of any surface effects from the final synthesis step and to prepare the surface prior to FIB milling, samples were cut out from the foil, mechanically ground (to about 1/3 of the total thickness from one side) with 1200 and 2000 grit size SiC paper, and finally polished with diamond suspensions (1 and 0.3 µm in particle size). The final average roughness (Ra) was ≈ 10 nm.

The microstructure of the Cu/Nb nanolaminate is shown in the TEM micrograph of Fig. 1(a). The average layer thickness was ≈ 63 nm, although there is a significant variability in the layer thickness due to processing method. The layers were about one grain thick and the in-plane grain size was on average more than 10 times longer than the individual layer thickness [26] (Fig. 1a). Therefore, the areal density of Cu–Cu and Nb–Nb grain boundaries was much smaller than the Cu-Nb interfaces, which control the deformation response. The interface structure of Cu/Nb MNLs manufactured by ARB shows a dominant orientation relationship $\{112\}<111>_{Cu} \parallel \{112\}<110>_{Nb}$ between Cu and Nb [4, 32], which only start to form as the individual layer thickness is reduced below the micron-scale. Furthermore, interfaces are faceted and contain misfit dislocations with both in-plane and out-of-plane Burgers vectors [33].

Tensile tests of the Cu/Nb MNLs were carried out in two different orientations (parallel and perpendicular to the layers) within a TEM (FEI Talos F200X) operated at 200 keV. The mechanical deformation was applied with a diamond flat punch using a Hysitron PI95 TEM PicoIndenter platform. The flat punch compresses a Push-to-Pull (PTP) loading device that transforms the motion of the flat punch into an opening of the central gap of the PTP device. Both ends of the tensile specimen were glued across the gap in the PTP device (Fig. 1b). The displacement rate of the flat punch of the PicoIndenter was set to 1 nm/s, which led to an average strain rate of 0.05 $s^{-1}$ in the specimen. The load and the flat punch displacement were measured and recorded in real time using the Performech® advanced control module integrated with the TriboScan™ software. The load carried by the specimen was determined from the applied load when the specimen was in the PTP device minus the load necessary to deform the PTP device up to the same displacement without the specimen, as indicated in [34]. In addition, real-time videos were recorded at 63 frames/s during the mechanical tests.

Dog-bone tensile samples were milled out from bulk Cu/Nb nanolaminates (Fig. 1c), using a dual-beam FIB/SEM system (Helios NanoLab 600i, FEI) following the procedure detailed in Fig.

S1 (supplementary material), with the gauge length parallel and perpendicular to the layers. The prepared specimens were then transferred onto the PTP device for mechanical testing using an micromanipulator (Easy Lift™). Both ends of tensile samples were carefully attached to the PTP device using ion beam-deposited Pt at a low current of 24 pA (Fig. 1d). In order to minimize the ion beam damage on the gauge section, only the parts outside of gage section were locally exposed to the ion beam during Pt deposition. The final dimensions of the central gauge of the dog-bone specimens were 2 μm (length) × 500 nm (width) × 120 nm (thickness).

The engineering stress was determined from the load applied to the specimen divided by the initial cross-section. The average engineering strain was computed from the displacement of the central gage section of the specimen in the loading direction, which was measured from video images using a digital image correlation software (Vic-2D).

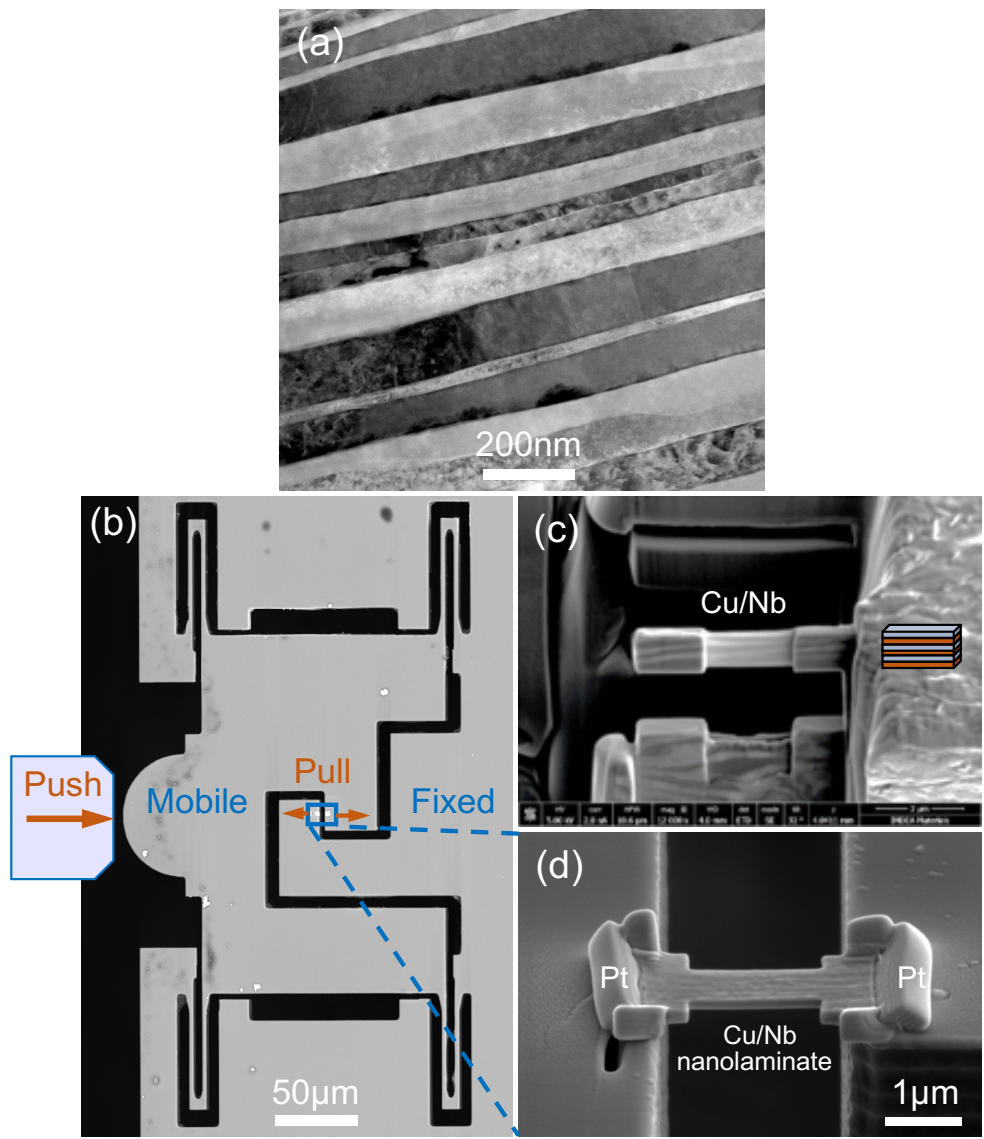

Fig. 1. (a) TEM-HAADF micrograph of the Cu/Nb nanolaminate, showing the nanolayered structure. (b) PTP device used to carry out the tensile tests within the TEM. The flat punch attached to the PicoIndenter compresses the PTP device, which transforms this displacement into an opening of the central gap where the tensile specimen is attached. (c) Tensile dog-bone specimen prepared by FIB with the layers parallel to the loading orientation. (d) Tensile specimen – attached to the PTP device with Pt – ready for testing within the TEM.

# 3. Results

## 3.1. Deformation parallel to the layers

The engineering stress-strain curve of the Cu/Nb MNL deformed parallel to the layers is depicted in Fig. 2. The initial response up to 0.62% was linear and no evidence of dislocation movement was found in the TEM images. A deformation pop-in occurred at this point, which was associated with a strain burst (see video 1 in the supplementary material) and to the elastic unloading of the specimen. Further straining led to the development of plastic deformation by CLS, which occurred in both Cu and Nb layers, as shown in Video 1 (supplementary material). An example of CLS of a single dislocation in Cu is shown in the TEM micrographs extracted from this video in Fig. 3. The dislocation travelled 163 nm in 64 s, leading to an average velocity of ~ 2.54 nm/s.

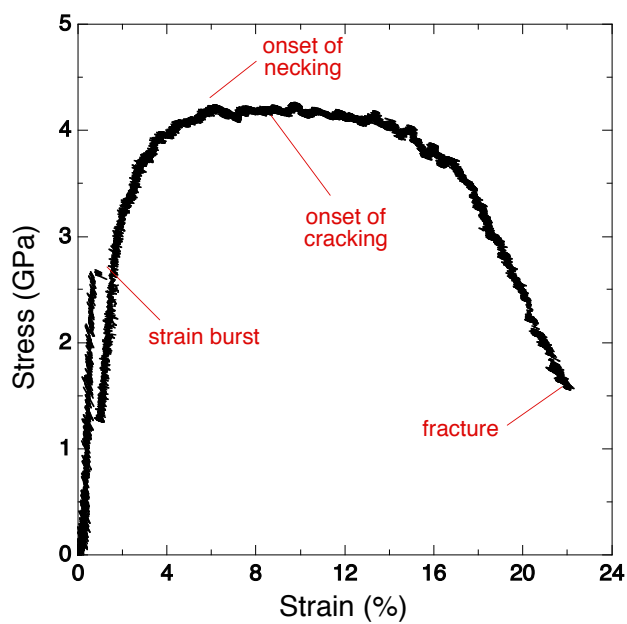

Fig. 2. Engineering stress-strain curve of the Cu/Nb nanolaminate deformed parallel to the layers.

Further plastic deformation was accommodated by CLS in both Cu and Nb layers, both co-deforming under nearly isostrain conditions. Gliding dislocations preferred to nucleate at the Cu/Nb interface, as shown in Fig. S2 (supplementary material). Deformation was homogeneous throughout the gage length up to a strain of 6%, at which the maximum stress was attained. Further straining led to the localization of the deformation in the center of the tensile specimen and to the propagation of a crack. This process is depicted in the TEM micrographs in Fig. 4, which were extracted from Video 1 (supplementary material). Localization of the deformation started in the two external Cu layers at both sides of the specimen (Fig. 4b). It led to the formation of a crack inclined at approximately 45º with respect to the loading axis in the Cu external layer at an applied strain of 8% (Fig. 4c). However, the crack orientation deviated when it reached the Nb layer and propagated in the second Cu layer in a direction perpendicular to the loading axis (fig. 4d). The deflected crack stopped at the Cu/Nb interface and further straining led to a large crack tip blunting in front of the second Nb layer. In parallel, another crack started to propagate from the right-hand side of the tensile specimen at a different height (Fig. 4d) and final failure took place by the linkage of both cracks through a shear band in the center of the specimen (fig. 4e).

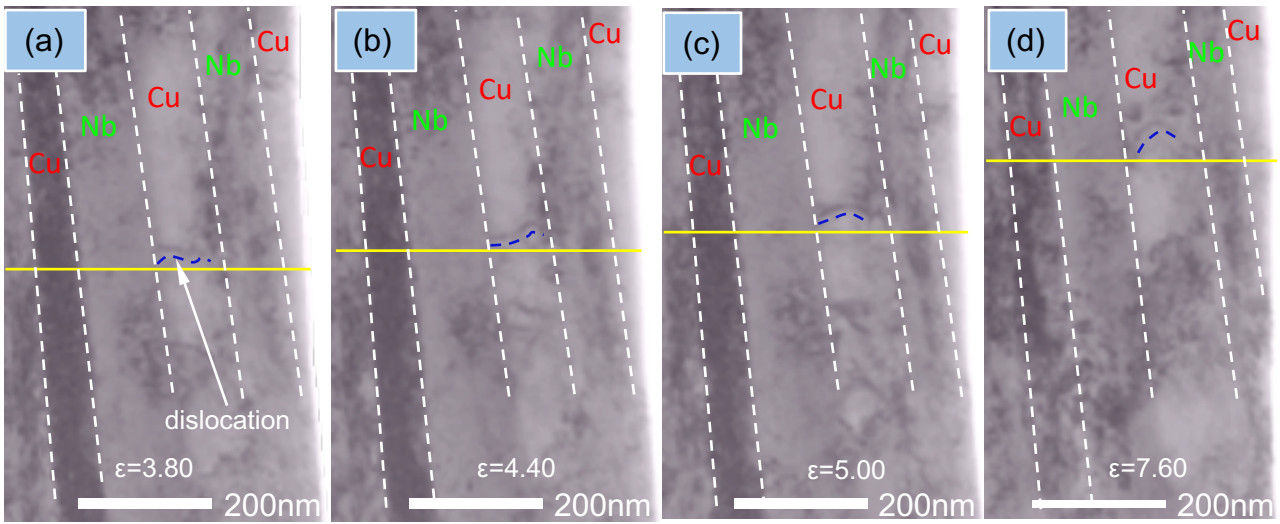

Fig. 3. (a) - (d) A series of TEM micrographs showing the motion of a single dislocation in a Cu layer by the CLS mechanism. The dislocation line is highlighted with a blue dashed curve for clarity. The single dislocation glided for ~ 163 nm in 64 s, leading to an average velocity of ~ 2.54 nm/s. The applied strain corresponding to each frame is indicated. Frames extracted from Video 1 (supplementary material).

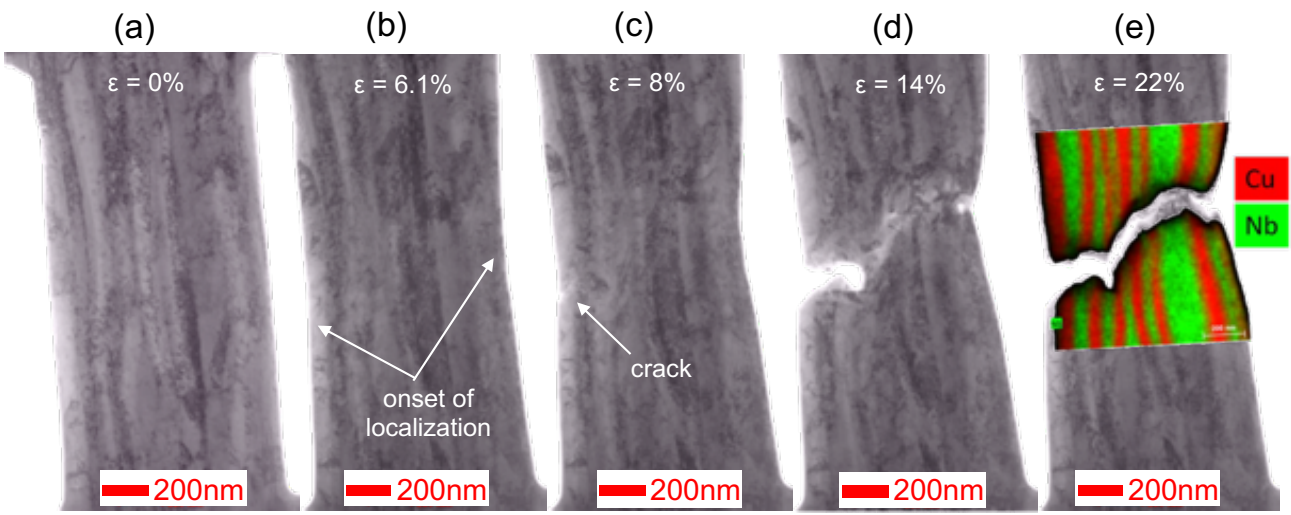

Fig. 4. Fracture mechanisms during deformation of the Cu/Nb MNL in the orientation parallel to the layers. (a) Initial microstructure. (b) onset of localization at the maximum applied stress. (c) Crack nucleation on the external Cu layer. (d) Crack propagation and blunting at the Cu/Nb interface. (d) Final crack path with the energy-dispersive X-ray microanalysis composition map overlaid. The applied strain corresponding to each frame is indicated. Images were extracted from video 1 (supplementary material).

*3.2. Deformation perpendicular to the layers*

The engineering stress-strain curve corresponding to the Cu/Nb NML deformed in the orientation perpendicular to the layers is plotted in Fig. 5. After the initial elastic region, a strain burst was also observed at an applied strain of 0.75%. However, the deformation remained elastic after the strain burst and no evidence of dislocation motion was observed in the TEM images until the applied strain reached 1.7%, which corresponds to an applied stress of ≈ 2 GPa. Plastic deformation by CLS was the dominant deformation mechanism from that strain onwards. The TEM micrographs in Fig. S3 (supplementary material) show that the thick Cu and Nb layers in this specimen encompassed several grains across the layer thickness but the grain boundaries were not strong obstacles to the dislocation motion. Deformation was rapidly localized in the thicker Cu layer situated near the bottom of the specimen, which was evidently softer (video 2 in the supplementary material). The fracture of the sample was brittle after the maximum stress was attained and the crack nucleation and propagation could not be resolved in the TEM images. Necking was only observed in the thick Cu layer (Fig. 5c). The crack path was within the Cu layer and the crack did not nucleate at the Cu/Nb interface (Fig. 5d).

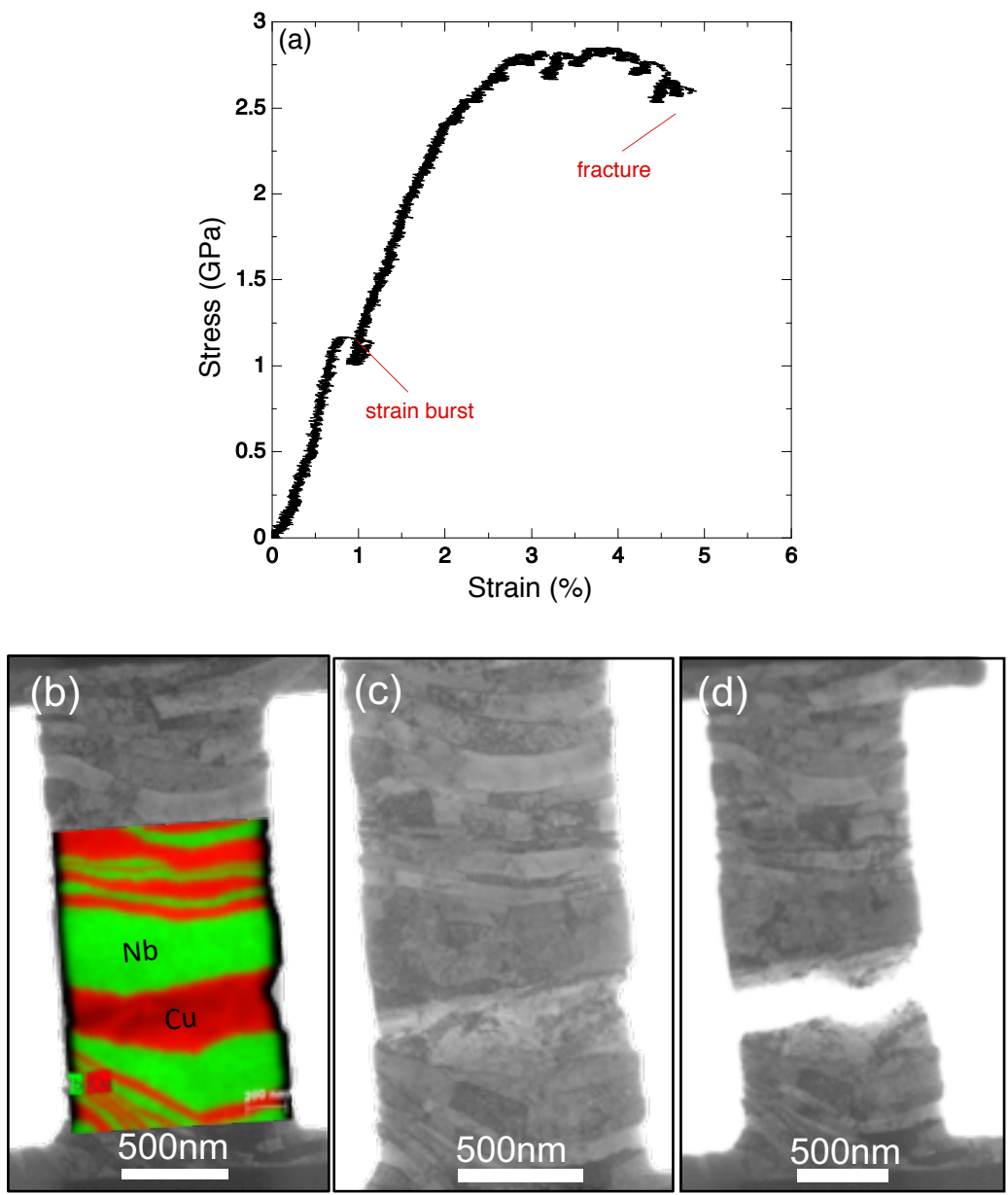

Fig. 5. (a) Engineering stress-strain curve of the Cu/Nb nanolaminate deformed perpendicular to the layers. (b) Bright field TEM micrograph of the specimen prior to deformation with the energy-dispersive X-ray microanalysis composition map overlaid. (c) and (d) Images extracted from the video 2 (Supplementary material) just before and after the sudden fracture throughout the Cu layer.

## 4. Discussion

The mechanisms of deformation and fracture of Cu/Nb MNL manufactured by ARB changed with the orientation, leading to huge anisotropy in the mechanical properties. The strength and the strain to failure in specimens deformed parallel to the layers were ~ 4.2 GPa and 22%, respectively, and they were reduced to ~ 2.8 GPa and 4.5% when deformation was perpendicular to the layers. These results are very different from those obtained in bulk specimens of ARB Cu/Nb with an average layer thickness of 63 nm [19], whose strength was ~ 0.90 GPa while the strain to failure was around ~ 7%.

The higher strength reported here for the ARB Cu/Nb MNs (as compared with the bulk specimens in [19]) could be attributed to the small thickness of the electron-transparent specimens, around 120 nm, tested inside the TEM in the present work. In this case, the number and length of dislocations sources is limited, leading to very large stresses (2.7 GPa and 1.2 GPa in the parallel and perpendicular orientations, respectively) to initiate plastic deformation, which began with a strain burst. This behavior is typical of very thin films in which plastic deformation is controlled by a limited number of dislocation sources [35]. After the onset of plasticity, plastic deformation by CLS was homogeneous in all the layers of the specimen deformed in the parallel orientation and the strain hardening rate was very high.

However, the specimen loaded perpendicular to the layers deformed elastically after the strain burst and plastic deformation by CLS was only evident when the applied stress reached 2 GPa. It started in one Cu layer whose thickness was much larger than that of all the other Cu layers and was eventually spread out to all the layers in the nanolaminate. Nevertheless, the dislocation activity was always higher in the Cu thicker layer, which was responsible for the limited strain hardening rate and the localization of damage at very low strains. As a result, the behaviour of the specimen tested in the perpendicular orientation was brittle.

On the contrary, the specimen deformed in the parallel orientation presented a very ductile behaviour because the process of damage localization and crack propagation (which was almost instantaneous in the specimen deformed in the perpendicular orientation) was hindered by the layered structure. More dislocation activity was also observed in the Cu layers, which is correlated to the greater number of slip systems available in the softer Cu layers. The observed extended plasticity led to localization of the plastic deformation in one external Cu layer causing the nucleation and development of a crack oriented at 45º with respect to the loading axis, which stopped at the Cu/Nb interface. Further propagation of the crack along the next Nb and Cu layers took place in the direction perpendicular to the loading axis and inhibited the formation of a shear crack, which was the expected failure mode taking into account that the samples were very thin (120 nm) and are in plane stress conditions. The crack stopped again at the next Cu/Nb interface and could not propagate across the Nb layer. Instead, significant crack blunting occurred, which was linked to the emission of dislocations at 45º from the crack tip. Finally, failure took place by the coalescence of this crack with another crack coming from the opposite side of the specimen in a different plane through a shear crack.

These observations indicate that damage tends to localize in the Cu layers in ARB Cu/Nb nanolaminates. This behaviour hinders the ductility in the specimens oriented in the perpendicular direction, particularly if there is one very thick Cu layer. On the contrary, damage in the Cu layers in the specimens deformed parallel to the layers cannot propagate easily throughout the multilayer structure because the Nb layers stop the crack. This leads to a damage tolerant behaviour and to a very large ductility in tension.

## 5. Conclusions

The mechanisms of deformation and fracture in Cu/Nb nanoscale multilayers manufactured by ARB were studied by means of tensile tests within a TEM in orientations parallel and perpendicular to the layers. The results obtained from these tests provided novel insights (that cannot be obtained otherwise) into the deformation and fracture processes in metallic nanolaminates. Plastic deformation by confined layer slip (CLS) was found in both orientations for Cu and Nb layers. In the case of deformation in the parallel direction, plastic deformation started with a strain burst and exhibited a high strain hardening rate, while plastic deformation was rapidly localized in a thicker (and, thus, softer) Cu layer in the perpendicular orientation. As a result, the strain hardening in this orientation was limited and failure was brittle, taking place by the sudden propagation of a crack in the thick Cu layer at an applied strain of 4.5%. No signs of interface debonding were observed in either of the orientations. In the parallel orientation, both Cu and Nb nanolayers deformed homogeneously up to an applied strain of 8%. Localization of the deformation in one of the external Cu layers led to the formation of a crack, whose propagation stopped at the Cu/Nb interfaces. Significant crack blunting and emission of dislocations took place before the crack could propagate through the Nb layers and failure took place by the coalescence of this crack with another crack coming from the opposite side in a different plane through a shear crack at 45º. As a result, the Cu/Nb nanolaminate deformed parallel to the layers presented a ductile behaviour with a failure strain of 22% and significant necking before fracture.

**Acknowledgements**

This investigation was supported by the European Research Council (ERC) under the European Union's Horizon 2020 research and innovation programme (Advanced Grant VIRMETAL, grant agreement No. 669141). Dr. Z. Liu would like to acknowledge the support from the Marie Sklodowska-Curie Individual Fellowship program through the project MINIMAL (grant agreement No. 749192). The authors are indebted to Dr. N. A. Mara and Dr. I. J. Beyerlein who provided the Cu/Nb nanolaminates.

**SUPPLEMENTARY MATERIAL**

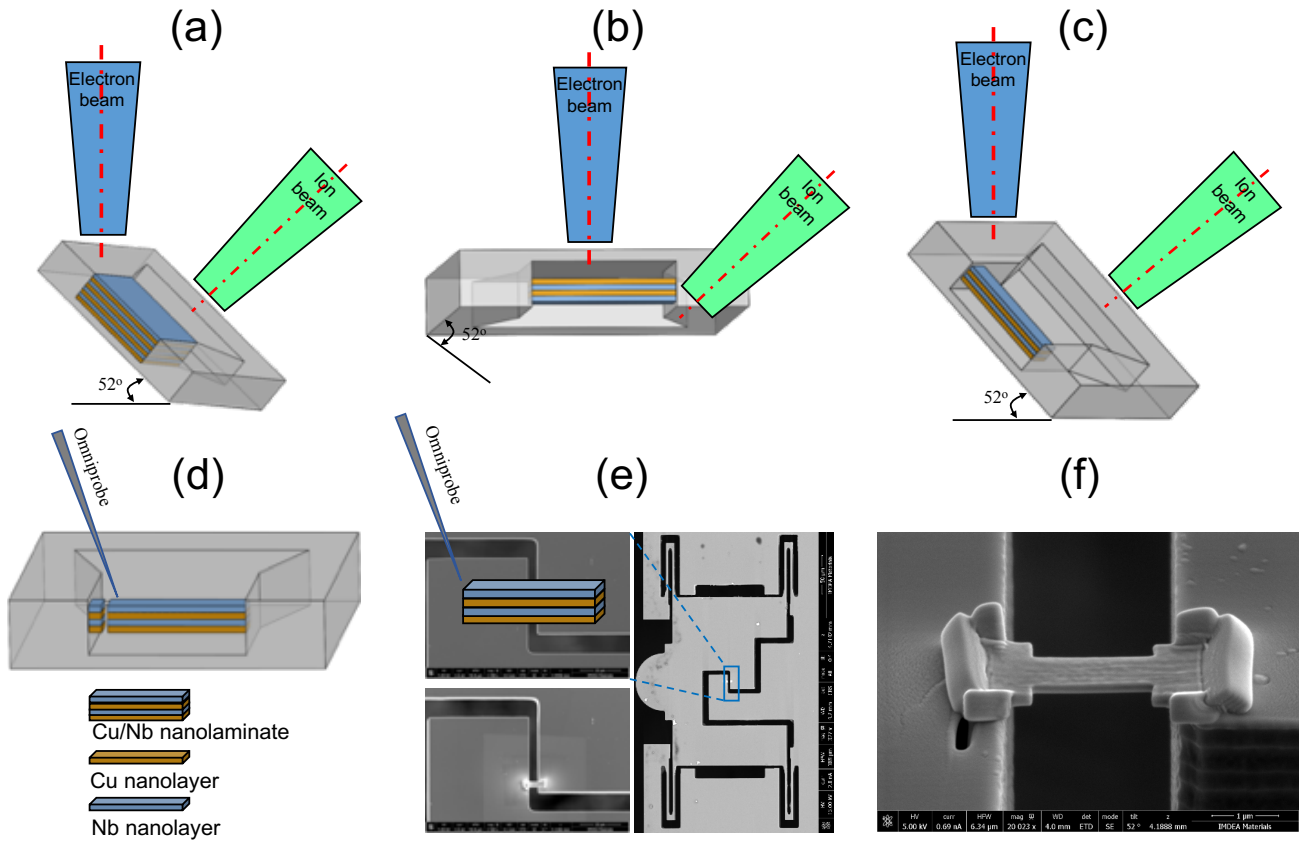

Fig. S1. Manufacturing of dog-bone shape specimens for *in situ* tensile tests within the TEM using focused-ion beam milling. (a) – (c) Milling a 3D thin parallelepipedic specimen from the bulk sample. (d) Lift-out the specimen from the bulk sample using an Omniprobe manipulator. (e) Transfer and fix of the specimen into the PTP device using FIB and Pt deposition. (f) Milling the parallelepipedic specimen into a dog-bone shape with FIB.

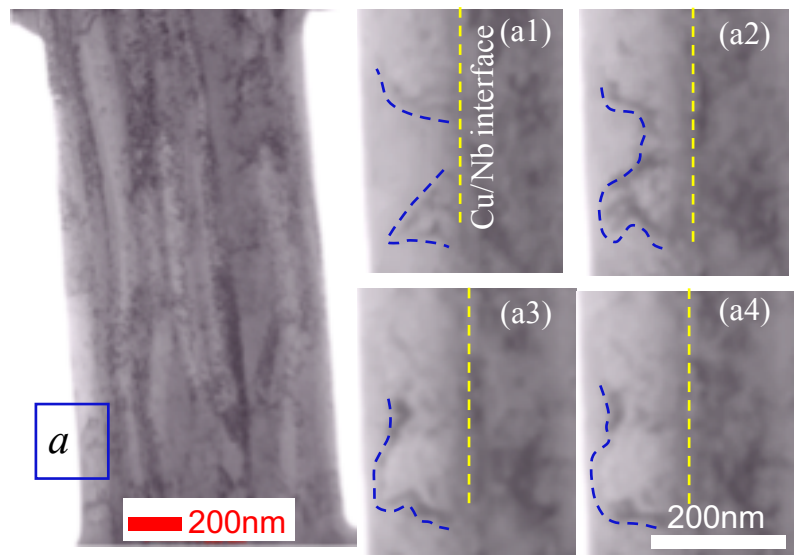

Fig. S2. Dislocation nucleation at the Cu/Nb interface at the point a. The frames, extracted from Video 1 (supplementary material), show one dislocation nucleated at Cu/Nb interface, which propagates in the Cu layer. Such phenomenon agrees well with the results reported in [22].

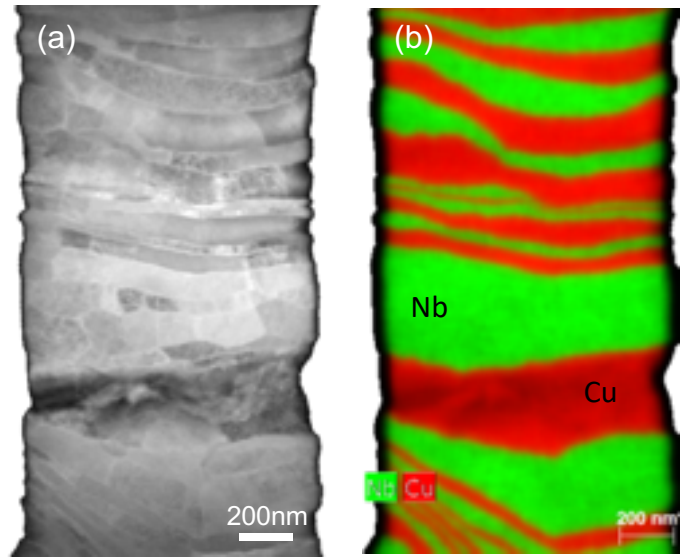

Fig. S3. (a) High-angle annular dark-field TEM micrograph of a Cu/Nb nanolaminate tensile sample. The thick Cu and Nb layers included several grains across the layer thickness. (b) high resolution energy-dispersive X-ray microanalysis composition map showing individual Cu and Nb nanolayers. Fracture occurred through the thick Cu layer.

Video 1. Tensile deformation of the Cu/Nb nanolaminate parallel to the layers.

Video 2. Tensile deformation of the Cu/Nb nanolaminate perpendicular to the layers.